\documentclass[aps,prl,reprint,superscriptaddress,amsmath,amssymb,preprintnumbers,showpacs]{revtex4-1}
\usepackage[normalem]{ulem}

\usepackage{enumerate}
\usepackage{graphicx}
\usepackage{color}
\usepackage{float} 
\usepackage{xspace}
\graphicspath{{./}}

\newcommand{\abs}[1]{\left| #1 \right|} 

\newcommand{\floqel}[1]{\xi_{E_\text{e}}}
\newcommand{\floqnuc}[1]{\xi_{E_\text{N}}}

\begin{document}
\title{Strong-Field Polarizability-Enhanced 
  Dissociative Ionization}
\author{Lun \surname{Yue}}
\affiliation{Institute for Physical Chemistry and Abbe Center for Photonics, Friedrich-Schiller University, 07743 Jena, Germany}
\author{Philipp \surname{Wustelt}}
\affiliation{Institute of Optics and Quantum Electronics, Friedrich-Schiller University, 07743 Jena, Germany}
\author{A. Max \surname{Sayler}}
\affiliation{Institute of Optics and Quantum Electronics, Friedrich-Schiller University, 07743 Jena, Germany}
\author{Gerhard G. \surname{Paulus}}
\affiliation{Institute of Optics and Quantum Electronics, Friedrich-Schiller University, 07743 Jena, Germany}
\author{Stefanie \surname{Gr\"{a}fe}}
\affiliation{Institute for Physical Chemistry and Abbe Center for Photonics, Friedrich-Schiller University, 07743 Jena, Germany}
\date{\today} 

\begin{abstract}


  We 
  investigate dissociative single and double ionization of HeH$^+$ 
  induced by intense femtosecond 
  laser pulses. 
  By employing a semi-classical model with nuclear trajectories moving on field-dressed surfaces and ionization events treated as stochastical jumps, 
  we identify a strong-field mechanism wherein the molecules dynamically align along the laser polarization axis and stretch towards a critical internuclear distance before getting dissociative ionized. 
As the tunnel-ionization rate is greater for larger internuclear distance and for aligned samples, ionization is enhanced. 
The strong dynamical rotation is traced back to a maximum in the parallel component of the internuclear-distance-dependent polarizability tensor. 
Qualitative agreement with our experimental observations is found. 
Finally the criteria for observing the isotope effect for the ion angular distribution 
is discussed.

\end{abstract}

\pacs{33.80.Wz, 33.80.Eh, 42.50.Hz}
\maketitle

The ionization and dissociation of small molecules in intense laser fields is of fundamental interest and has captured the attention of physicists for many years \cite{Giusti-Suzor95,Posthumus04,Heide2018}.
When the ratio of the laser frequency to the peak electric field is sufficiently small,
the ionization process can be considered as an electron tunneling through the instantaneous barrier formed by the field and the Coulomb potential of the system \cite{Keldysh1964}.
In molecules, such tunnel-ionization rates depend on the spatial separation between the nuclei \cite{Zuo1995,Seideman1995,Plummer1996,Posthumus1998,Kamta2005}, as well as the molecular orientation with respect to the laser polarization axis, where the ionization rate follows the shape of the highest-occupied molecular orbital \cite{Alnaser2004,Tong2005,Tolstikhin2011}. 
In addition to these fixed-nuclei properties, the molecules will dynamically rotate and stretch in the field, potentially leading to fragmentation \cite{Posthumus1998}.
Here, we theoretically identify a new fragmentation pathway that involves the combination of the aforementioned strong-field dynamics.
Namely, due to the force resulting from the internuclear-distance-dependent polarizability tensor, the molecule is simultaneously aligned and stretched towards a specific internuclear distance in the field-dressed ground state before being ionized. 
We denote it as polarizability-enhanced dissociative ionization (PEDI) and support our findings with experimental data.

Polarizability effects have been explored extensively in strong-field physics,
e.g. it has regularly appeared in the interpretation of strong-field ionization experiments \cite{Holmegaard2010,Pfeiffer2011}, and the anisotropic polarizability is often exploited in molecular alignment experiments \cite{Stapelfeldt2003,Damari2016,Lin2016}.
In dissociative ionization studies involving short laser pulses (tens of femtoseconds (fs) duration), often only the vibrational degrees of freedom are considered, while the rotational dynamics are disregarded. This is based on the intuition that the field-free rotational timescale of picoseconds is much greater than the vibrational timescale of fs and thus rotational motion can be safely neglected. However, as several works employing semiclassical methods \cite{Dietrich1993,Plummer1996,Tong2005b} have shown, rotational dynamics are crucial for the understanding of the angular distribution of the final ion fragments.
Even at
lower intensities where ionization is negligible and pure dissociation is the dominating fragmentation process, 
it was shown
that molecular rotations play a role \cite{Anis2008,McKenna2008,Anis2009,Ursrey2012,Yu2016} 
for pulses as short as $\sim 5$ fs.



\begin{figure}
  \includegraphics[width=0.48\textwidth]{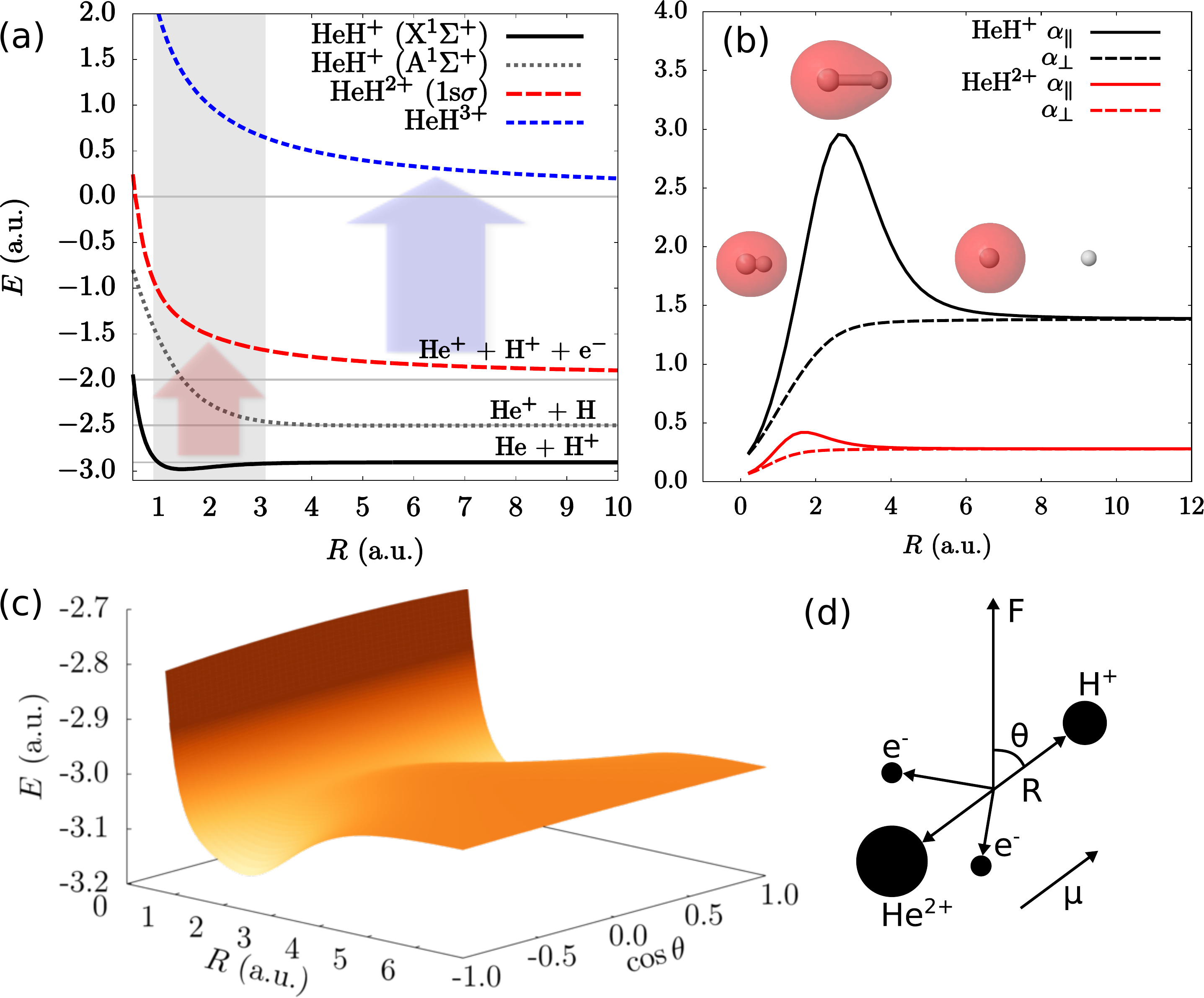}
  \caption{
 (a) Field-free BO curves considered in this work for HeH$^+$ and its daughter ions. The Franck-Condon region from the initial vibrational distribution in $\text{HeH}^+$ is indicated by the shaded region. (b) Components of the 
polarizability tensors as a function of internuclear distance. The densities (isosurface value 0.04) of HeH$^+$ (X$^1\Sigma^+$) are plotted for $R=$1.0, 2.6 and 6.0. (c) Cycle-averaged field-dressed potential for $F=0.53$ [see Eq.~\eqref{eq:1} and the following text]. (d) Sketch of HeH$^+$. \label{fig:1}
  }
\end{figure}

For our studies, 
we choose to focus on the simplest stable polar molecule, HeH$^+$, sketched in Fig.~\ref{fig:1}(d). Aside from HeH$^+$ being a reference
 system, our choice is motivated by the fact that the two lowest electronic states, X$^1\Sigma^+$ and A$^1\Sigma^+$ [see Fig.~\ref{fig:1}(a)], have a large energy separation of $13-39$ eV in the Franck-Condon region 
and a vanishing dipole coupling between them at large $R$.
This highly limits the effect of the excited states and leads
to the essential physics occuring in the electronic ground state. This is in stark contrast to H$_2^+$, where the two lowest 
charge-resonance 
states \cite{Mulliken1939} are energetically separated by a few IR photons at intermediate $R$, 
and degenerate and strongly coupled at large $R$. 
Indeed, most of the prominent breakup processes first discovered in H$_2^+$ 
depend on the efficient population of the first excited state. These processes include above-threshold dissociation \cite{Giusti-Suzor1990,Jolicard1992,Yue13,Lu2017}, bond-hardening and bond-softening \cite{Bandrauk1981,Bucksbaum1990,Zavriyev1993,Frasinski1999}, 
electron localization \cite{Seideman1995,Kling2006,Staudte2007,Waitz2016,Xu2017},
above-threshold Coulomb explosion \cite{Esry2006} and charge-resonance enhanced ionization \cite{Zuo1995}. 
Although PEDI should be present in H$_2^+$ as well as in other more complicated molecules, more prominent population and coupling of excited states would lead to many processes being intertwined, making an identification of PEDI difficult. 
It should be mentioned that the first excited state in $\text{HeH}^+$ has received attention 
 in terms of the enhanced ionization (EI) phenomenon in polar molecules \cite{Kamta2005,Dehghanian2013}: at a critical internuclear distance, due to the crossing of the two lowest Stark-shifted energy levels, enhanced population and ionization of the first excited state occurs. 
However, the EI discovery was based on a fixed-nuclei picture, and we find for a more realistic scenario, i.e. with moving nuclei and a molecule which is initially prepared close to the equilibrium $R$, the dominating effect is that ionization is enhanced due to the joined dynamics of rotation and stretching of the molecule. Recent quantum calculations with vibrating $\text{HeH}^+$ also confirm the dominance of ionization from the ground state \cite{Wang2017}.



In this letter, strong-field dissociative single and double ionization (SI and DI) of HeH$^+$ is simulated employing a semiclassical approach with classical nuclear trajectories moving on field-dressed surfaces and ionization treated as stochastical jumps. Such an approach allows us to treat rotation, stretching, single and double ionization, while at the same time keep the computational effort manageable to perform scans over an extensive laser-parameter space with inclusion of the initial rotational temperature, vibrational- and intensity-focal-volume-averaging effects. A full quantum treatment at the laser intensities considered in this work ($\lesssim10^{16}\; \text{W/cm}^2$) 
is computationally prohibitive.

In the $\text{HeH}^+$ sketch in Fig.~\ref{fig:1}(d), $R$ denotes the internuclear distance and $\theta$ the angle between the field polarization and the molecular axes.
The field-dressed energy surfaces read (atomic units are used unless stated otherwise)
\begin{equation}
  \label{eq:1}
  \begin{aligned}
    E^{(s)}&(R,\theta,t)
    ={E}_0^{(s)}(R)-\mu^{(s)}(R)F(t)\cos\theta\\
    &-\tfrac{1}{2}F^2(t)    \left\{\alpha^{(s)}_{\perp}(R)+\left[\alpha^{(s)}_{\|}(R)-\alpha^{(s)}_{\perp}(R) \right]\cos^2\theta  \right\},
  \end{aligned}
\end{equation}
with the superscript $(s)$ indicating either the ground states of HeH$^+$ and HeH$^{2+}$, or the Coulomb potential of HeH$^{3+}$ [Fig.~\ref{fig:1}(a)], 
$F$ the instantaneous electric field, $E_0^{(s)}$ the Born-Oppenheimer curves, $\mu^{(s)}$ the permanent dipole moments and $\alpha_\perp^{(s)}$ ($\alpha^{(s)}_\|$) the molecular-frame perpendicular (parallel) components of the polarizability tensors (obtained on the CASSCF(15/2)/aug-cc-pVQZ \cite{Roos1980} level of theory calculated with MOLCAS \cite{molcas8}). 
For HeH$^+$ (X$^1\Sigma^+$), the anisotropic polarizability $\alpha_\|^{(s)}-\alpha_\perp^{(s)}$ exhibits a distinct maximum at $R_{c}=2.6$ [Fig.~\ref{fig:1}(b)], which leads to distinct wells at $R_c$ and $\cos\theta=\pm 1$ in the cycle-averaged field-dressed energy surface [Fig.~\ref{fig:1}(c)]. 
The peak in $\alpha^{(s)}_\|$ is understood in terms of the plotted density isosurfaces in Fig.~\ref{fig:1}(b): For $R\rightarrow 0$ ($R \rightarrow \infty$), the electron cloud is spatially confined and resides almost completely on the united-atom $^5$Li nucleus ($^4$He nucleus), resulting in small polarizabilities,
while at intermediate $R$'s the density is spatially extended, resulting in large polarizabilities. 
Recent quantum chemistry calculations on alkali dimers have also noticed the polarizability maximum and its possible implications for alignment experiments \cite{Deiglmayr2008}.

\begin{figure}
  \centering
  \includegraphics[width=0.48\textwidth]{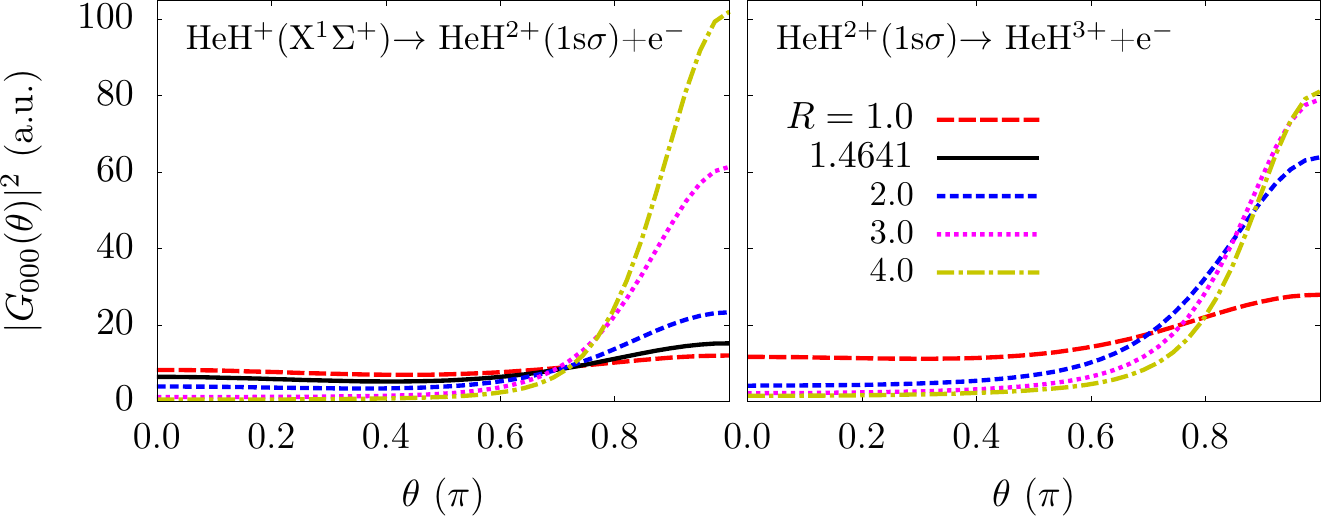}
  \caption{\label{fig:2}Calculated structure factors for single and (sequential) double ionization $\abs{G^{(s)}(\theta;R)}^2$ included in the ME-WFAT ionization rates.}
\end{figure}

Each trajectory moves classically on the instantaneous field-dressed surface [Eq.~\eqref{eq:1}], with ionization treated as stochastical jumps \cite{Dietrich1993,Fiedlschuster2016}. For the pulses considered in this work, the Keldysh parameter is $\gamma<1$, indicating the tunneling regime of strong-field ionization \cite{Keldysh1964}. We employ lowest-order many-electron weak-field-asymptotic theory (ME-WFAT) \cite{Tolstikhin2014}, where the ionization rate $\Gamma^{(s)}(F,\theta;R)=\abs{G^{(s)}(\theta;R)}^2W^{(s)}(F;R)$ is given in terms of the \textit{structure factor} $\abs{G^{(s)}(\theta;R)}^2$ and the \textit{field factor} 
\begin{equation}
  \label{eq:2}
  W^{(s)}=\frac{\kappa^{(s)}}{2} \left( \frac{4{\kappa^{(s)}}^2}{F} \right)^{2Z^{(s)}/\kappa^{(s)}-1}\exp\left(-\frac{2{\kappa^{(s)}}^3}{3F}\right),
\end{equation}
with $\kappa^{(s)}=\sqrt{2I_P^{(s)}}$, $Z^{(s)}$ the nuclear charge seen by the outgoing electron asymptotically and $I_P^{(s)}$ the $R$-dependent ionization potentials. An additional emperical factor $\exp[-14{Z^{(s)}}^2F/{\kappa^{(s)}}^5 ]$ is applied to counteract the overestimation of the rates at large $F$ \cite{Tong2005}. The structure factors for $R=1-4$ are obtained using 
the method in Ref.~\cite{Yue2017} 
and extrapolated to the values of the relevant atoms for $R\rightarrow\infty$. As shown in Fig.~\ref{fig:2}, the electron favourably tunnels from the hydrogen side, in agreement with earlier works \cite{Kamta2005,Kamta2007,Tolstikhin2011}. 
The initial conditions of $R$ and $p_R$ for each trajectory are chosen randomly according to a Husimi distribution of a given vibrational state, and the molecules are chosen to be randomly oriented. The initial angular momenta $p_\theta$ are distributed according to a Boltzmann-distribution to account for the high experimental rotational temperature in the ion source, $T=3400\pm 300\ \text{K}$ \cite{Loreau2011,Ketterle1985}. 
The laser wavelength is $800\ \text{nm}$ and the pulse has a $\sin^2$ envelope with the full width at half maximum $\tau=34\ \text{fs}$.
For each intensity and initial vibrational state, $10^3$ trajectories are released. 
All the results presented in this work are averaged over the initial vibrational-state distribution \cite{Loreau2011}, the laser beam focal volume \cite{Wang2005}, and the carrier-envelope phase, unless states otherwise.

The experimental setup is identical to that described in Ref.~\cite{Wustelt2018pre}. Briefly, the HeH$^+$ ions are synthesized in a duoplasmatron ion source, accelerated to 10 keV kinetic energy, and focused towards the laser interaction region, where a tabletop Ti:Sapphire laser system provides 
800nm-pulses with peak intensities of up to $10^{17}\ \text{W/cm}^2$ and intensity full width at half maximum duration $\tau\sim 34$ fs.




\begin{figure}
  \centering
  \includegraphics[width=0.48\textwidth]{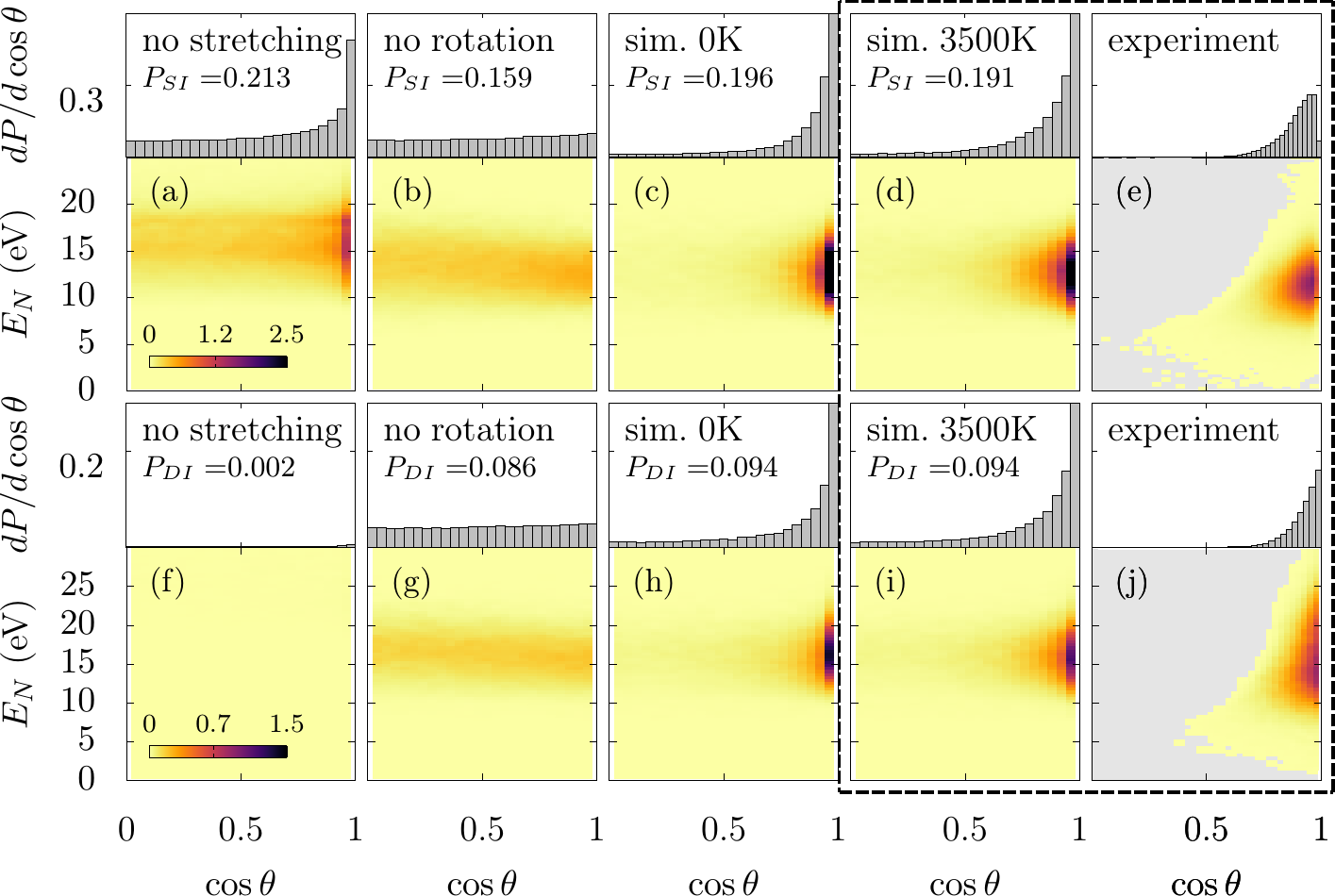}
  \caption{\label{fig:3} KER- and orientation-dependent fragmentation yields from single (upper panels) and double dissociative ionization (lower panels) of $\text{HeH}^+$ for $\lambda=800$ nm, $I=9\times 10^{15}$ W/cm$^2$ and $\tau=34$ fs. 
    Panels (e) and (j): experimental results at laser peak intensity ($\sim 10^{17}$ W/cm$^2$), with the shaded area denoting absolute zero counts due to the experimental setup (see text). Colorscale in arbitrary linear units.
    }
\end{figure}

Figure \ref{fig:3} shows the kinetic energy release (KER) and angular distributions of the nuclear fragments from SI and DI. Due to the 
high number of optical cycles, 
the directional ionization asymmetry for $\theta=0$ and $\theta=\pi$ [Fig.~\ref{fig:2}] are averaged out, 
resulting in symmetric yields in the intervals $\cos\theta\in[0,1]$ and $[-1,0]$, of which we only consider the former.

\begin{figure}
  \centering
  \includegraphics[width=0.48\textwidth]{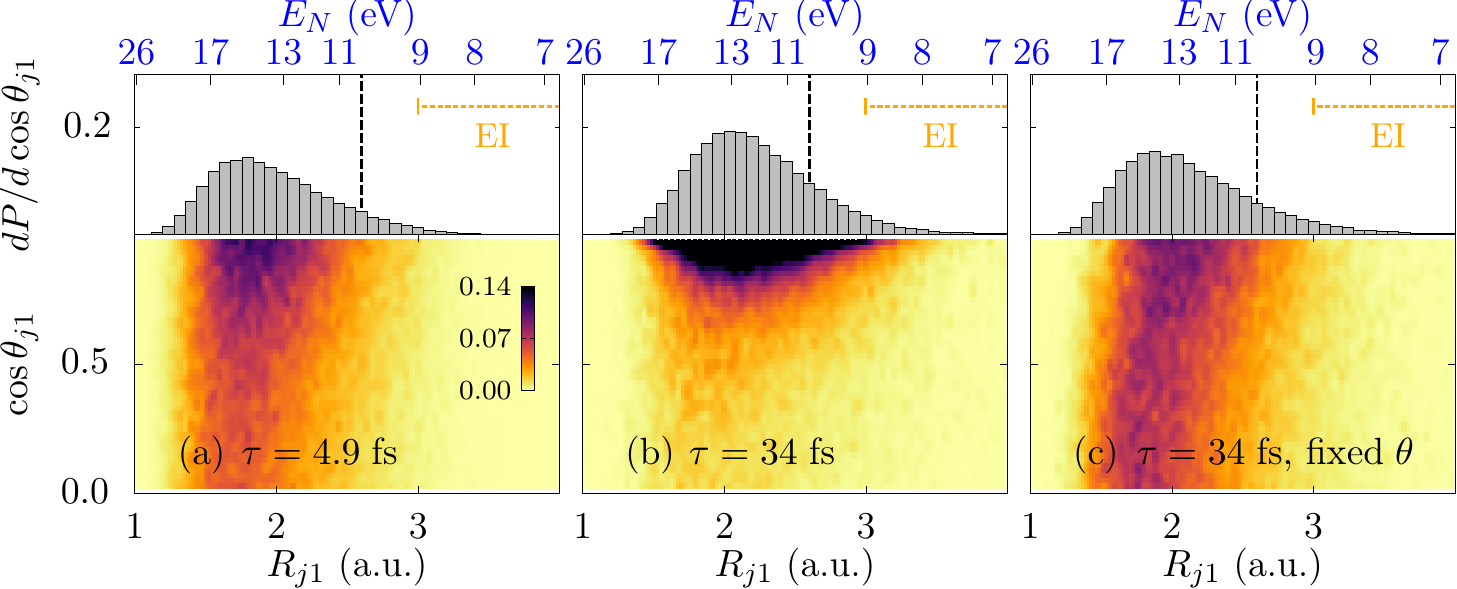}
  \caption{\label{fig:4} Simulated distribution of $R$ and $\cos\theta$ for the singly-ionized molecules at the instant of ionization for $I=9\times 10^{15}\ \text{W/cm}^2$ and $T=0\ \text{K}$. (a) $\tau=4.9\ \text{fs}$; (b) $\tau=34 \ \text{fs}$; (c) fixed $\theta$ (disabled rotation) with $\tau=34\ \text{fs}$. The vertical dashed line indicates $R_c$, the horizontal dotted line approximately traces the enhanced ionization region from \cite{Dehghanian2013} (not included in our simulations), and
 the and the upper (blue) axis shows the kinetic energy by the reflection principle.
    }
\end{figure}

The SI experimental result in Fig.~\ref{fig:3}(e) shows that the fragments have KERs of 7-17 eV with the angular distribution aligned along the laser polarization direction ($\cos\theta=1$).
 Due to the Faraday cup used in the experimental setup to block the non-fragmented HeH$^+$ beam, the yields at $\cos\theta\approx 1$ are not fully detected. 
Also, due to the finite size of the detector, the measureable range of $\cos\theta$ is limited, resulting in zero counts in the grayed region in Figs.~\ref{fig:3}(e). 
Our simulation results with $T=3500\ \text{K}$ in Fig.~\ref{fig:3}(d) agrees with the experiment, with the fragments aligned along the laser polarization axis and having KER 7-18 eV, corresponding to ionization at $R\sim 1.5-3.5$ (see Fig.~\ref{fig:4}). Compared with Fig.~\ref{fig:3}(c), a nonzero temperature is seen to broaden the angular distribution. Three intertwined effects lead to the final aligned ion distribution: dynamic alignment during the laser pulse, geometric alignment \cite{Posthumus1998,Anis2009} denoting the orientation-dependent ionization (Fig.~\ref{fig:2}), and post-ionization alignment \cite{Tong2005b}. 
By artificially disabling the molecular rotation in the simulations, we see in Fig.~\ref{fig:3}(b) that the angular distribution is quasi-flat and cannot mimick the experimental results, thus geometric alignment plays a lesser role for the final ion angular distribution.
Simulation results where we disabled the molecular stretching shown in Fig.~\ref{fig:3}(a) have too high KER (peaked at 17 eV) compared to the full result in Fig.~\ref{fig:3}(d), corresponding to ionization events occuring at smaller $R$'s.

The measured DI yields in Fig.~\ref{fig:3}(j) are also aligned, with KER 8-25 eV, corresponding to the second ionization event occuring between $R\sim 5-25$ \cite{Wustelt2018pre}. The simulation with $T=3500\ \text{K}$ in Fig.~\ref{fig:3}(i) agrees with these observations. The angular distributions for SI and DI are similar, because after a molecule is singly ionized, it will dissociate and its moment of inertia $MR^2$ ($M$ the reduced mass) increases, effectively freezing $\theta$ before DI. 
For both SI and DI, the ionization probabilities $P_{SI}=0.192$ [Fig.~\ref{fig:3}(d)] and $P_{DI}=0.071$ [Fig.~\ref{fig:3}(i)] are larger when the molecules are allowed to strech and rotate,
indicating increased ionization due to dynamical alignment and stretching towards $R=R_c$.


We estimate the rotational timescale as the time it takes a trajectory to reach from half of the well depth in Fig.~\ref{fig:1}(c) to the potential minima, i.e.
$\tau_\text{rot}=\sqrt{M}h(R)/F $, with $h(R)\approx2.62R/\sqrt{\alpha_\|-\alpha_\perp}$. At $R_c$, with 
 $M_{\text{HeH}^+}=1469$ and $F=0.37$, we have $\tau_\text{rot}=12.8\ \text{fs}$ which is comparable to the field-free vibrational period $\tau_\text{vib}=11.5\ \text{fs}$ and 37
 times shorter than that of the field-free rotational period of 478 fs. 
Dynamical rotation is expected to be prominent when the rotational timescale $\tau_\text{rot}$ is comparable to or shorter than the pulse duration $\tau$, which is the case for $\tau=34\ \text{fs}$ used in the experiment. 

Fig.~\ref{fig:4} shows the $(R_{j_1},\theta_{j_1})$-distribution of the singly-ionized molecules at the instant of ionization. For the 
results in Fig.~\ref{fig:4}(c) with rotation artificially disabled, the ionization 
occur over a broad range of $\cos\theta_{j_1}$.
With increasing $\cos\theta_{j_1}$, the ionization 
are seen to be shifted towards larger $R_{j_1}$, e.g. we have $R_{j_1}\approx 1.8$ for $\cos\theta_{j_1}=0$, and $R_{j_1}\approx 2.2$ for $\cos\theta_{j_1}=1$. 
Also, more ionization events occur for the aligned molecules ($\cos\theta\approx 1)$. 
The reason is clear: without rotation, the aligned molecules need to stretch towards larger $R$ [see Fig.~\ref{fig:1}(c)] 
 where the lower $I_p$ and the favourable structure factor in Fig.~\ref{fig:2} 
leads to more ionization.
The projected $R_{j1}$-distribution in the upper panel of Fig.~\ref{fig:4}(c) peaks at $R_{j_1}\approx 1.9$, corresponding to $E_N\approx 14\;\text{eV}$ by reflection. 
This is in disagreement with the experimental KER maximum in Fig.~\ref{fig:3}(e) at $E_N\approx 11\;\text{eV}$. 
 For the simulation results in Fig.~\ref{fig:4}(b) where rotational motion is included, most molecules ionize after they are dynamically aligned, with the projected $R_{j_1}$-distribution peaked at $R\approx2.1$, corresponding to a reflected KER $E_N\approx 12.5\ \text{eV}$, in 
better
agreement with the experimental value.
For the short pulse $\tau=4.9 \ \text{fs} \lesssim \tau_\text{rot}$ in Fig.~\ref{fig:4}(a), the molecules do not have much time to rotate and stretch, resulting in a smaller $R_{j_1}$'s and thus much higher KERs.
We mention that we have not taken the EI effect \cite{Kamta2005,Kamta2007,Dehghanian2013} into account in our ionization rates, which will take place at larger $R$ (horizontal dashed lines in Fig.~\ref{fig:4}). 
Since most ionization events occur before this region, we expect 
its effect to be small.


\begin{figure}
  \centering
  \includegraphics[width=0.48\textwidth]{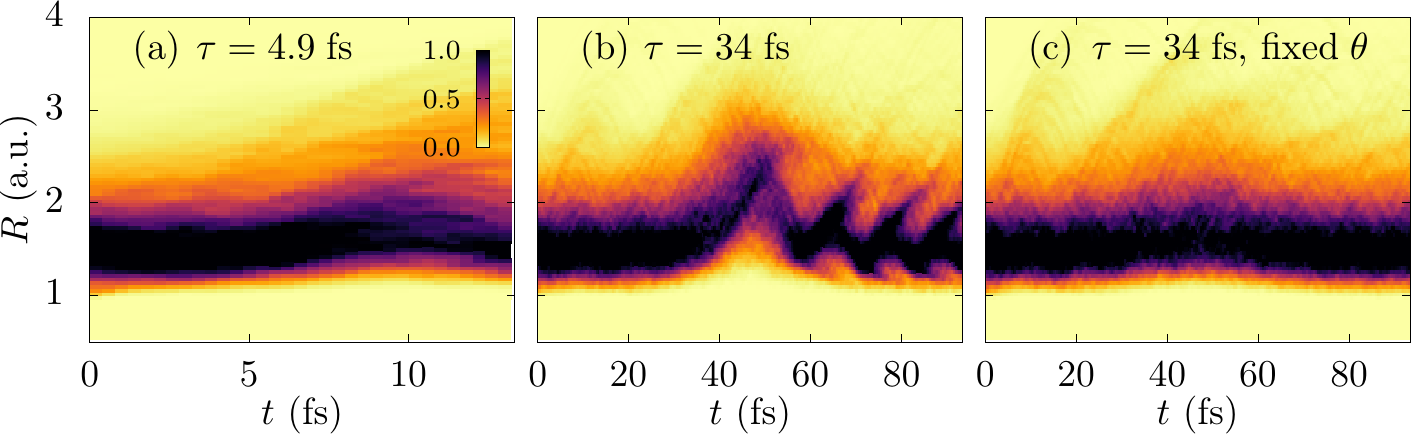}
  \caption{\label{fig:5}
    Simulated time-evolution of the nuclear densities in the artificial case where ionization is turned off, for $I=5\times 10^{15}\ \text{W/cm}^2$ and $T=0\ \text{K}$. (a) $\tau=4.9\ \text{fs}$; (b) $\tau=34 \ \text{fs}$; (c) fixed $\theta$ (disabled rotation) with $\tau=34\ \text{fs}$. The results are vibrational averaged, but not intensity-focal-volume averaged.
    }
\end{figure}


To understand more in detail the bound-state dynamics before ionization, we have considered the artificial case with ionization switched off. The time-evolution of the bound nuclear densities are shown in Fig.~\ref{fig:5}. Since the low-intensity regions in the focal-volume dominate, we have only considered a single intensity $I=5\times10^{15}\ \text{W/cm}^2$ without performing the focal-volume averaging. For the short pulse in Fig.~\ref{fig:5}(a), $\tau<\tau_\text{rot}\approx\tau_\text{vib}$, and the trajectories do not have time to rotate and vibrate, which results in a minimal change of the density during the pulse. 
This is different for a long pulse shown in Fig.~\ref{fig:5}(b). The molecules have time to align and stretch towards $R_c$, and at the field maximum $t=46.7\ \text{fs}$ a substantial part of the density has reached $R\in[2,3]$. During the second half of the pulse, the already aligned molecules contract back towards the equilibrium $R_0$, with density oscillations due to the vibrational motion distinctly observed in Fig.~\ref{fig:5}(b). When the molecule is not allowed to rotate [Fig.~\ref{fig:5}(c)], similar to the case of a short pulse, the stretching of the nuclei follows the dressed potential, but only molecules initially aligned along the laser polarization direction ($\cos\theta=1$) stretch towards $R_c$.

\begin{figure}
  \centering
  \includegraphics[width=0.48\textwidth]{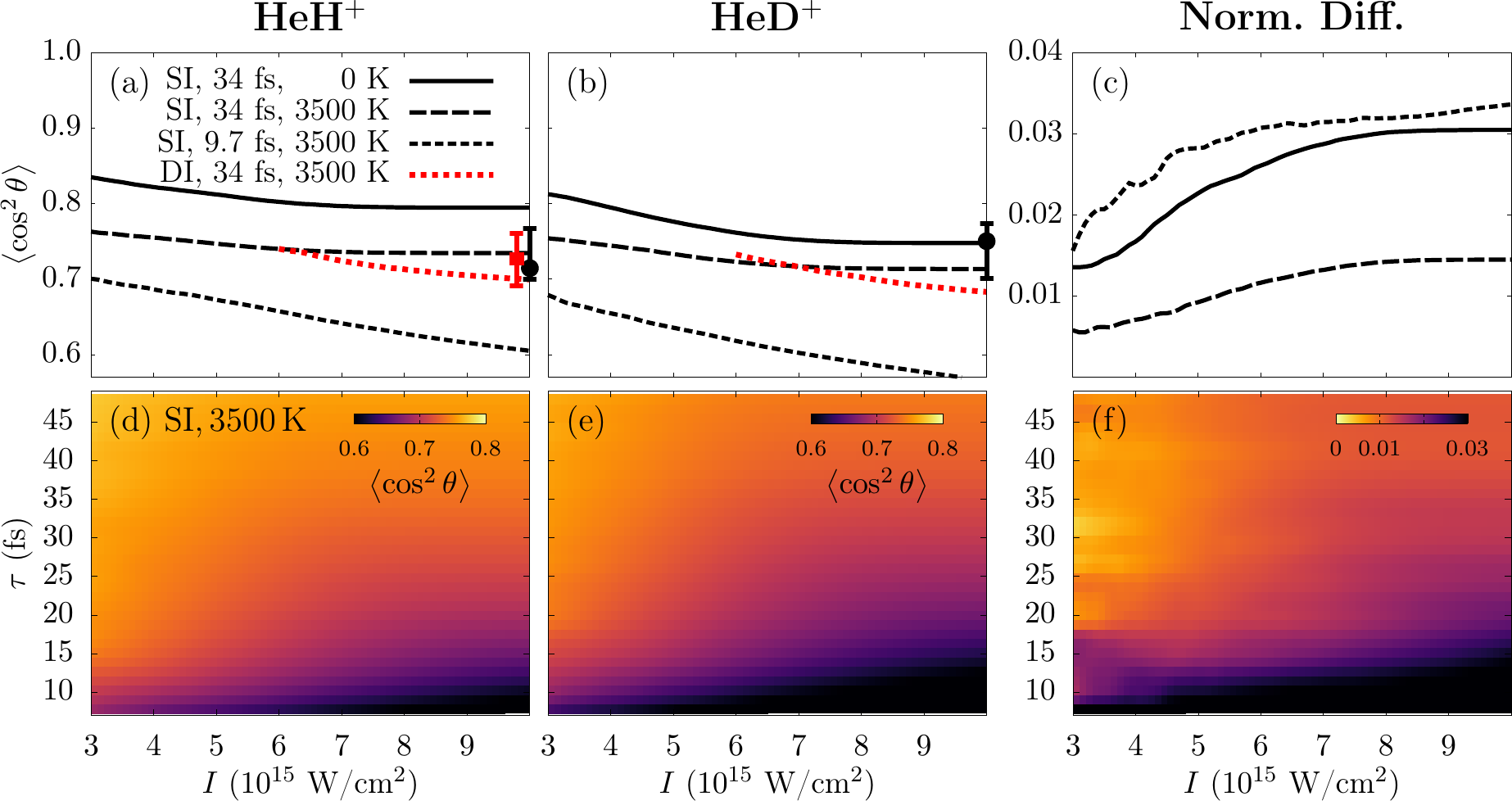}
  \caption{\label{fig:6} Alignment parameter $\left<\cos^2\theta\right>$ for (a,d) HeH$^+$, (b,e) HeD$^+$, and (c,f) their normalized difference. In (a,b), the fitted experimental results at peak intensity, $800\;\text{nm}$ and $34\;\text{fs}$ for SI and DI are given by the (black) circle and (red) square, with the bars denoting the uncertainty in the fit (see text).
  }
\end{figure}
 
The degree of alignment can be characterized by the alignment parameter $\left<\cos^2\theta\right>$ (equal to 1 for completely aligned samples and $\frac{1}{3}$ for randomly oriented samples) and is given as a function of $I$ for HeH$^+$ in Fig.~\ref{fig:6}(a). For SI, higher temperature and shorter pulses result in broader distributions. 
With increasing intensities, the detected fragments are seen to be less aligned, 
despite the rotational timescale $\tau_\text{rot}\propto 1/F$ being shorter. 
This is indeed a characteristic of dynamical rotation: with increasing $I$, ionization will occur earlier during the pulse, leaving the molecule with less time to rotate \cite{Posthumus1998}.
 The experimental values for $\left<\cos^2\theta\right>$ are shown in Fig.~\ref{fig:6}(a) as a circle and square for SI and DI, respectively. 
To avoid inaccuracies resulting from the properties of the experimental setup
we consider only the counts in the interval $E_N\in[0,E_\text{max}]$ in Figs.~\ref{fig:3}(e) and \ref{fig:3}(j), with $E_\text{max}=7\ \text{eV}$, and fitted the missing yields at $\cos\theta\approx 1$ with a function $f(\cos\theta)=a\cos^2\theta+b\cos^c\theta$. The bars belonging to the data points are for the extremal values of $\left<\cos^2\theta\right>$ for $E_\text{max}\in\text{[4, 9] eV}$ 
and 
are seen to be consistent with the simulation results.

Since $\tau_\text{rot}\propto\sqrt{M}$, the dynamical rotation is 
influenced 
by the specific isotope. Indeed, recently the isotope effect in tunneling ionization of molecules has attracted considerable attention \cite{Tolstikhin2013,Wang2016}. We finally investigate whether an isotope effect can be observed in the ion angular distributions of $\text{HeH}^+$ and $\text{HeD}^+$.
Fig.~\ref{fig:6}(b) presents the simulated and experimental results for $\text{HeD}^+$. With the employed experimental laser parameters, however, this isotope effect cannot be conclusively resolved. To experimentally identify this would require shorter pulses, as depicted
in Fig.~\ref{fig:6}(c) showing the normalized difference between the curves in Figs.~\ref{fig:6}(a) and \ref{fig:6}(b), $S\equiv \left(\left<\cos^2\theta\right>_{\text{HeH}^+}-\left<\cos^2\theta\right>_{\text{HeD}^+}\right)/ \left( \left<\cos^2\theta\right>_{\text{HeH}^+}\right. \\  \left. +\left<\cos^2\theta\right>_{\text{HeD}^+}  \right)$.
For the $34\;\text{fs}$, 3500 K case, $S<0.014$, while for the $9.7\;\text{fs}$ pulse $S$ has more than doubled to 0.033. 
In Figs.~\ref{fig:6}(d)-(f) we show $\left<\cos^2\theta\right>$ for $\text{HeH}^+$, $\text{HeD}^+$ and their relative differences over an extensive pulse regime. It is seen that the isotope effect for dynamical rotation is pronounced for pulse durations of less than $10\ \text{fs}$, potentially allowing for an experimental detection.

In conclusion, 
we have identified polarizability-enhanced dissociative ionization, a strong-field molecular breakup pathway where the molecules dynamically align and stretch towards a specific internuclear distance before ionization, with geometric alignment playing a lesser role. For our studies we have focused on the fundamental polar molecules $\text{HeH}^+$ and $\text{HeD}^+$, and we believe that the effect is quite general and more pronounced for polar diatomics. Indeed, the maximum in the anisotropy polarizability is present in all diatomics except for odd-charged molecular ions where the parallel polarizability will monotonically increase with $R$ \cite{Mulliken1939}. 




\acknowledgments
The authors acknowledge support from the German Research Foundation (DFG-SPP-1840 ``Quantum Dynamics in Tailored Intense Fields''). L. Y. thanks Johannes Steinmetzer for aiding in the calculation of the molecular polarizabilities.


%

\end{document}